\documentclass{svproc}

\usepackage{url}
\usepackage{graphicx}
\usepackage[boxruled,vlined,linesnumbered]{algorithm2e}

\begin{document}
\mainmatter              
\title{Testing of Smart TV Applications: Key Ingredients, Challenges and
Proposed Solutions}

%
%
\author{Bestoun S. Ahmed* \and Miroslav Bures}
\institute{Department of Computer Science, Faculty of Electrical Engineering\\
Czech Technical University\\
Karlovo nám. 13, 121 35 Praha 2, Czech Republic\\
\email{albeybes@fel.cvut.cz}\\
\email{buresm3@fel.cvut.cz}
}

\maketitle              

\begin{abstract}
Smart TV applications are software applications that have been designed to run on smart TVs which are televisions with integrated Internet features. Nowadays, the smart TVs are going to dominate the television market, and the number of connected TVs is growing exponentially. This growth is accompanied by the increase of consumers and the use of smart TV applications that drive these devices. Due to the increasing demand for smart TV applications especially with the rise of the Internet of Things (IoT) services, it is essential to building an application with a certain level of quality. Despite the analogy between the smart TV and mobile apps, testing smart TV applications is different in many aspects due to the different nature of user interaction and development environment. To develop the field and formulate the concepts of smart TV application testing, this paper aims to provide the essential ingredients, solutions, answers to the most critical questions, and open problems. In addition, we offer initial results and proof of concepts for a creeper algorithm to detect essential views of the applications. This paper serves as an effort to report the key ingredients and challenges of the smart TV application testing systematically to the research community.
\keywords{Smart TV application testing, Software Testing, Model-based Testing, Internet of Things (IoT)}
\end{abstract}
\section{Introduction}

A connected TV, which popularly called smart TV, is a technological assemblage device among computer and traditional television. The device is a combination of conventional TV terminal, operating system (OS), and digital contents in which all of them are connected to the Internet. Smart TVs are providing different digital services like multimedia, gaming, Internet browsing, on-demand entertainment access, a various online interactive session in addition to broadcasting media. In fact, these devices were expected to be more intelligent, interactive, and useful in the future \cite{Jung2011}. Recently, the electronic companies along with IT firms were rising investments in the technological advancements of these devices by launching new terminals and applications for smart TVs. It is expected shortly that these devices will be a frequent part of our smart homes within an Internet of Things (IoT) context\footnote{https://read.bi/2L4CDSI}. This explains why the smart TV market worth \$265 Billion by 2016\footnote{https://bit.ly/2HxnMkL}.

Just like the new technological smart devices, smart TVs are operated by an OS with different applications (apps) installed on it. Although the OS is the key software for operation, the installed apps on the smart TV brings different uses and functionalities to the device. At a glance, the smart TV app may look like a mobile app due to the similarities of the OSs or the development kits. Due to this \textquotedblleft fake\textquotedblright{} similarity, one may think of testing smart TV apps just like the mobile app testing. However, in fact, testing smart TV apps is different due to the nature of user interaction with the app itself.

In mobile apps, the user is interacting with the device touchscreen (i.e., the application) directly by hand whereas, within smart TVs, the user is interacting with the app through another device which is the remote controller. Of course, some vendors are providing interaction by touchscreen to the users, but the way that application behaves is still based on the remote control device when it comes to testing practices. In addition, the user of any TV (including the smart TVs) is usually staying away from the screen and almost use the remote device to operate the apps all the time.

In the literature, mobile apps testing is well-studied, and many research directions have been established, (e.g., \cite{Zein2016,Sahinoglu2015,AMALFITANO20131}). However, testing smart TV apps is a new area and many challenges still without a solution, and many research questions may arise without answers. To address these challenges and questions, it is essential to explore the app structures, interaction ways, development environments, and the technology behind the apps. In doing so, this paper examines the key ingredients of smart TV app testing. The paper aims to address the most demandable questions. The paper also discusses the challenges addressed so far in the literature and open problems for test automation and generation. Based on that, a systematic framework for testing applications on Smart TVs is illustrated throughout a prototype. The framework includes the testing process, its steps, and also the test generation strategy. This will help to validate the different aspects of the applications before release. This could also serve as an initiative topic for further research in the near future. The framework will help to address and formulate more open problems and research questions. 

The rest of this paper is organized as follows. Section \ref{sec:Related-Works} summarizes the related works in the literature and those efforts in smart TV app testing that could be useful here. Section \ref{sec:Smart-TV-Apps-Technology} explains the technology behind the smart TV apps. Section \ref{sec:The-Analogy-and-Differences} illustrates some analogy and differences between mobile and smart TV apps. Section \ref{sec:Navigation-and-Control} describes the navigation and control mechanism of smart TV apps. Section \ref{sec:OpenProblems} discusses the open research problems in the smart TV app testing. Section \ref{sec:Towards-a-Systematic-testing} defines a prototype for a systematic automated testing strategy. Section \ref{FunctionaltestingSection} discusses the functional and non-functional testing Opportunities in Smart TV Applications. Finally, Section \ref{sec:Conclusion} give concluding remarks and also future research recommendations.

\section{Motivation and Literature\label{sec:Related-Works}}

Testing software applications on smart devices is considered to be a development and an evolution of testing practice from the traditional user interfaces (UI) like graphical user interface (GUI) and web application testing. The testing practices for these UIs have been studied extensively in the last decade, and as a result, many sophisticated methods, algorithms, and tools have been developed. Banerjee \textit{et al.} \cite{Banerjee:2013:GUI} studied more than 230 articles published between 1991-2013 in the area of GUI testing and Li \textit{et al.} \cite{Li:2014:TDW} surveyed the literature in two decades of web application testing. 

Mobile application testing could be considered as the first effort towards smart application testing. There are many differences between mobile apps and graphical/web UI. In fact, the main issue that makes the difference in the testing process is the user interaction with the application. In the standard GUI and web applications, the keyboard and mouse combination is still the standard user input to interact with the applications. However, this is not the case for mobile apps as the user interacts with the device touchscreen by fingers and hence, there would be different interaction behavior from various users. Although this issue leads to develop new testing strategies for mobile apps, still many of these strategies are taking benefits, wholly or partially, from the earlier methods and practices published for GUI and web application testing. For example, Amalfitano \textit{et al.} \cite{Amalfitano2015MobiGuitar} developed MobiGUITAR strategy for systematic mobile application testing from the GUITAR strategy \cite{Nguyen2014:Guitar} for GUI testing. An extensive study on mobile application testing is presented in \cite{Zein2016}.

Smart TV application is a new smart device application type. The views of the application are not like other applications. The application structure looks like web application as it relies on HTML, CSS, and JavaScript; however, the user interaction with the application differs from other types of applications. Usually, the user is not interacting with the application directly by hand, and it should be through another input device, which is the remote device. This could lead to think that the testing process is similar to the GUI or web application. However, the remote device does not behave like the standard mouse. While the standard mouse input device can move in every direction on the application, the remote device movement is restricted to four explicit directions. The interaction difference makes many obstacles and difficulties when it comes to testing process. While the general concepts of model-based testing are applicable here, the construction of the model and the model type makes the difference. For example, due to the different interaction nature, Nguyen \textit{et al.} \cite{Nguyen2014:Guitar} used Event Flow Graph (EFG) as a model of the GUI testing, whereas Amalfitano \textit{et al.} \cite{Amalfitano2015MobiGuitar} uses state machine as a model for the mobile application testing. In smart TV app, both EFG and state machine models are not applicable. In Smart TV app, each transition from a state to another is practically just one step, while this is not the case in other applications. For example, in the mobile app, the distance between two icons (states) does not make sense in the transition, while this is very important in the smart TV application, and that will lead to a different model. An important effort to formulate this model is done recently by Cui \textit{et al.} \cite{Cui2017}. Here, the Hierarchical State Transition Matrix (HSTM) is proposed as a model for the Android smart TV applications. While the model is promising, there is a need to develop and formulate it for the complex structure of different applications.

In fact, testing smart TV apps could be seen from different angles. For example, usability testing is one of the critical testing issues to address the interaction between the user and the smart TV through remote device. This will help to improve the quality of the user interfaces of the applications. Ingrosso \textit{et al.} \cite{Ingrosso2015} addressed this issue by using several users to test an e-commerce application on smart TV. Security testing is also an essential issue in the smart TV apps. However, we could not find a published study addressing security in Smart TV apps. Recently, Sabina C. \cite{Sabina2016} discussed and described some of the testing platforms for Smart TV apps. The study chooses Opera and Samsung TV Stores for testing the applications. The testing process relies on the upload of the applications to the Opera and Samsung application stores to verify them based on the code writing. Hence, there is no definition of the testing strategy itself, and that could not be considered as a formal testing process. The study has also addressed the importance of functional testing of these applications without giving details since it is a bachelor study with limitations.

Although it is essential from the industrial point of view, we could not find many companies giving solutions for smart TV apps testing. One of the exciting projects so far is the suite.st framework\footnote{https://suite.st}. The framework depends on record and replay testing style by using two different devices, one for recording the actions, and the other is for acting like an emulator. In fact, the platform dealing with the application just like a web application and uses record and replay style of testing being employed by SeleniumHQ\footnote{http://www.seleniumhq.org/}. The framework is a good startup for the industry to adapt selenium style of testing for smart TV apps. Although the framework claims that it is dealing with the functional testing of mobile apps, still the pass/fail criteria are not clear from an academic point of view. As a result, there is a need to define a test oracle for the framework. In addition, the framework does not rely on some automatic test generator for fully testing of the applications. In fact, defining a test oracle for smart TV application could be a new research direction as we will address it later in this paper.

\section{Smart TV Apps Development and Technology\label{sec:Smart-TV-Apps-Technology}}

Just like Android apps, smart TV apps are developed using Software Development Kits (SDK). The new versions of Android SDK supporting the development of smart TV apps. However, these applications can be run on Android Smart TV devices only. In fact, few SDKs were available for cross-platform development. For example, Joshfire\footnote{https://www.joshfire.com/} Smart TV SDK was a platform to develop applications to work on Google and Samsung TV devices but not on LG TV devices. Mautilus\footnote{https://www.mautilus.com} Smart TV SDK is also a platform for development, but still, the application is working on some versions of devices only. Smart TV Alliance\footnote{http://www.smarttv-alliance.org} was the most advanced SDK by supporting different features and platforms. However, the project is shut down, and the SDK is not available for download. 

Samsung Tizen SDK provides a set of tools and frameworks to develop smart TV apps through Tizen Studio. The SDK is depending on the latest web technologies such as JavaScript, CSS, HTML5, and W3C widget packaging. In fact, Samsung has established Tizen.Net which is a new cross-platform application development that has been integrated with Visual Studio.

Nowadays, most of the SDK tools are relying on a unified approach to the development technology for smart TV apps. The technologies behind the applications are JavaScript, HTML5, and CSS3. JavaScript is used as a standard programming language to program the behavior of the applications. The use of JavaScript adds the page jumping capability of the application. It enables the developer also to code complex expressions and calculations like conditional branches, and loops. The fifth version of the Hypertext Markup Language (HTML5) is used as the latest version for developing the web elements' structure and content. The HTML5 is essential to develop the structure of the application page even without the JavaScript code, but that will lack the interactivity with the user \cite{HTML5Book}. Finally, the third version of the Cascading Style Sheets (CSS3) is used for the presentation of these web elements and polishing them for better visualization. These essential components are forming the latest and best technology of the smart TV application, and also they are the newest technology for the World Wide Web. 

In general, Smart TV app could be one of two types, installed or cloud-based. Installed TV app is a stand-alone app installed on the smart TV without the need for the Internet connection, while the cloud-based TV app works as an interface between the cloud and the TV with a shallow content (almost no additional functionality) when there is no Internet connection.

\section{The Analogy and Differences of Smart TV and Mobile Apps\label{sec:The-Analogy-and-Differences}}

There are many similarities and differences between the Mobile and Smart TV apps. These similarities and differences could be seen in three dimensions, (1) Functionality, (2) Design, and (3) User interaction. 

Both applications are working on smart devices. Hence, the functionality could be similar, as they are both connected to the Internet. The mobile apps could be useful even without connection to the Internet; however, several smart TV apps are useless without the network connection. The computation power of the smart device also could define the functionalities of the application itself. In fact, the mobile apps could be more functional than smart TV apps because the mobile devices nowadays may have more computational power than smart TVs. In addition, the aim of the mobile apps is almost different from the smart TV apps. 

Speaking about the application design, there are many differences. For example, the size of the screen and icons could define the layout of the application. Smart TV screens are wider than the mobile devices. The background color of the smart TV apps could be different from the color in the mobile devices. From the user interaction point of view, smart TV apps are having less text entry as it is difficult to enter text from the remote device. Most of the smart TV apps are designed to get the content from the Internet when connecting whereas this is not the case for the mobile apps, as they could be standalone applications without Internet connections interfaces\footnote{https://bit.ly/2IiNb30}. The typical smart TV application is much more straightforward than the mobile app, especially in the design layout. 

The way that the user interacts with the application defines an essential difference between the smart TV and mobile apps. The user of the mobile app interacts directly with the application without an intermediate device, while in the smart TV application, the user interacts with the help of a remote device. In fact, the UI of the smart TV apps sometimes called 10-foot user interfaces since the 10 feet (3m) distance from the TV is the standard distance between the user and the TV. The developers are considering this distance when developing the user interface \cite{Sabina2016}. Using the remote device with this distance is not user-friendly and not responsive. Hence, the UI must consider this significant difficulty. As mentioned previously in Section \ref{sec:Related-Works}, this interaction difference will be significant also when approaching the testing process with model-based testing.

\section{Navigation and Control in Smart TV Apps\label{sec:Navigation-and-Control}}

\begin{figure}
\centering
\includegraphics[scale=0.2]{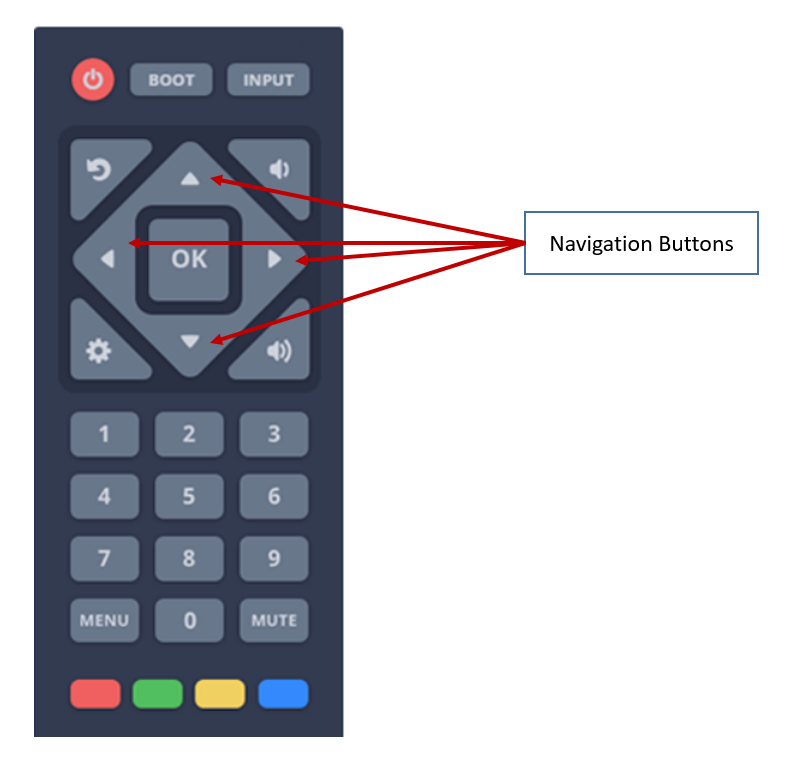}
\caption{\label{fig:TV-remote-controller}TV remote device}
\end{figure}

As mentioned previously, navigation on a smart TV application is through the remote device. Although some new TV devices are offering the direct interaction by the user with the screen, the most common interaction with the TV is still the remote device. The remote device consists of four essential navigation \textit{Right}, \textit{Left}, \textit{Up} and \textit{Down}. In addition, the remote device has an \textit{OK} button to choose any selected view on the application after exploration. These five key buttons should work properly while using an application. Figure \ref{fig:TV-remote-controller} shows an example of the TV remote device. 

In addition to those five buttons, there are many other buttons on the remote device that vary from a TV brand to another depending on the level of functionalities. Some of them are related to the hardware functionalities of the TV itself, as the power button to turn ON/OFF the TV. There are also ten buttons (from 0-9) for channel jumps and even entering numbers in text fields if necessary. 

\begin{figure*}
\begin{centering}
\includegraphics[scale=0.5]{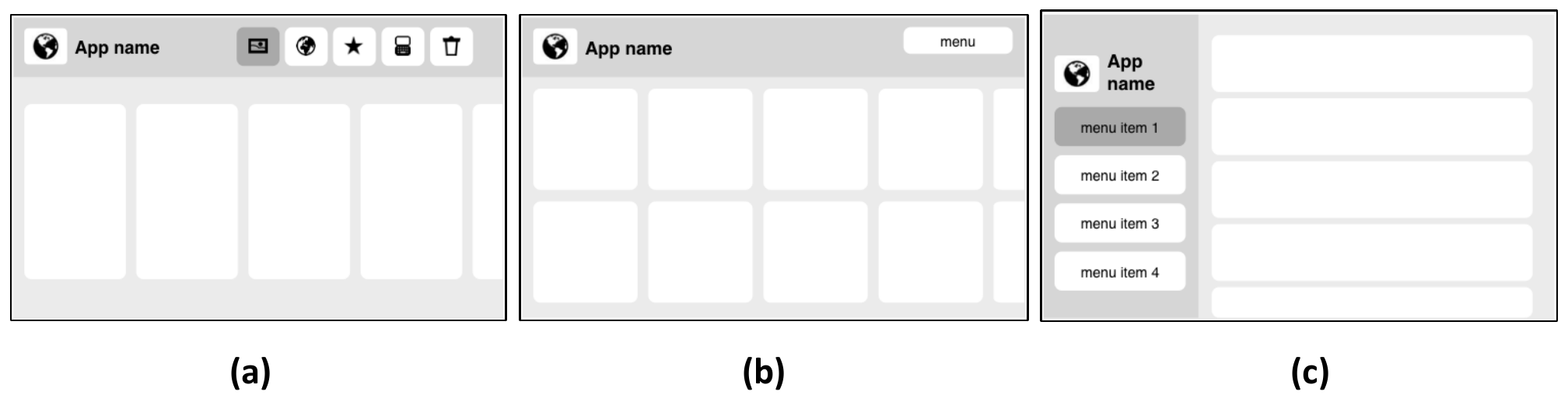}
\par\end{centering}
\caption{\label{fig:Three-main-layout} Three main layout design patterns for smart TV apps \cite{OperahDesignDevelope} }

\end{figure*}

The UI layout of any application plays a primary rule in the testing process. Understanding the layout could lead to an efficient test generator and runner. Smart TV apps are following some limited number of layout patterns. Figure \ref{fig:Three-main-layout} shows three main patterns in which most of the smart TV apps are following. In fact, layout (b) is mostly used, since it puts many views in one window.

The remote device is putting constraints on the navigation from a view to another because it supports just one step navigation. Hence, each move on the layout is a step. This would not be a problem when two views are adjacent; however, for those non-adjacent views, more than one step is needed to move from one view to another. This navigation is very important when coming to the test generation strategy based on the application's model. 

\section{Open Problems and Challenges\label{sec:OpenProblems}}

In this section, we discuss different problems and challenges that need to be addressed for the smart TV app testing. In the following subsections, we will address each problem, the challenges to solve the problem and our suggestions.

\subsection{Start Point of Navigation}

One of the first problems that the tester face when testing a smart TV app is the position of the navigational cursor. Technically speaking, from a JavaScript developer point of view, this happened when the focus point is not set in the application. For several applications on the store, this focus point is not set by the developers. As a result, when the application runs on the emulator, there is no pre-selected view on the application. The user must use the remote device to chose a view. Hence, the starting point of the navigator is missing. This problem is happening clearly with cloud-based TV apps because the views are changing in real-time with the cloud content. In fact, this is a challenging issue because it prevents the pre-generation of test sets. 

One solution to this problem is to let the tester choose the starting point of the testing. Yet, there could be a problem of good or bad selection point. Some starting points may lead to explore the app window sooner by navigating faster on the views.

\subsection{Repository and Benchmark}

In general, any software testing verification and validation process should be evaluated through some benchmarks. These benchmarks could be real instrumented programs with some properties for testing. For example, many testing strategies are using the benchmarks available at Software-artifact Infrastructure Repository website\footnote{http://sir.unl.edu/portal/index.php} for benchmarking and evaluation. For android testing, there are different applications for testing. For instance many papers were using \textit{TippyTipper}\footnote{https://tinyurl.com/yd77qfzd} , \textit{PasswordMaker Pro}\footnote{https://tinyurl.com/ma65bc8}, \textit{MunchLife}, \textit{K-9 Mail}\footnote{https://tinyurl.com/6mzfdaa}, \textit{Tomdroid}\footnote{https://launchpad.net/tomdroid}, \textit{AardDict}\footnote{https://github.com/aarddict/android/issues/44}, and a few other applications for testing. 

In smart TV apps testing, we don't have enough applications for benchmarking, and we don't have a repository to store some benchmarks. In fact, there are two reasons behind this. First, smart TV apps are new  and more time may be needed for the developers to create and publish open source applications. Second, the testing process of smart TV app is not defined yet, and the research is not initialized, in which this paper could be an effort toward that. Samsung maintains a page with some simple applications and examples\footnote{https://bit.ly/2qC5ncS}.

One solution for this difficulty is to develop applications for testing purposes. Here, the reliability of the testing process would be an issue. However, for better reliability, the testing and development groups could be separated. 

\subsection{Test Generator}

In mobile app testing, most of the test generation strategies were almost inspired by other UI test generation strategies. For example, the test generator strategy of MobiGUITAR\cite{Amalfitano2015MobiGuitar} framework was adapted from the GUITAR\cite{Nguyen2014:Guitar} framework for GUI testing. However, this method could not be followed in smart TV apps. Due to the user interaction difference in smart TV app, it is hard to adapt some test generator strategy from GUI or mobile app testing. For this reason, there is a need to develop a new test generation strategy. 

Although relying on previously investigating strategies is not clear at this early stage, following principles and concepts of model-based testing is still valid. Here, after deciding on the model and notations, the coverage criteria of the testing strategy would be another issue. Defining the coverage criteria depends mainly on the tested functional and non-functional requirements.

\subsection{Activity Exploration}

The test generation stage cannot be performed without input to the generator algorithm. For functional or non-functional testing, most probably, the input would be two things, the number of events to test and the coverage criteria. As mentioned previously, the coverage criteria can be defined based on a predefined testing strategy. However, getting the input views for the test generation algorithm may need an exploration of the entire UI activity (i.e., window) of the smart TV app. 

Activity exploration is not a big issue (at least technically) when we have the source code of the application, i.e., white box testing. A simple code crawler could scan the HTML5 and CSS3 files and detect the views by parsing the code, and then feed the generator algorithm by these views. However, catching the views in the testing process without having the source code (i.e., black-box testing) could be a tricky job. In fact, there is a need for a special algorithm due to the special interaction with the application by the remote device. 

In Section \ref{sec:Application-Creeper}, we will introduce an algorithm to creep the significant views of the application activity in a black-box manner.

\subsection{Stopping Criteria}

Stopping criteria in the smart TV app could be an issue, especially for the cloud-based applications. In the installed TV app, there is a finite number of views in which the creeper can catch them, and the testing strategy can cover. When this coverage criteria are met, the testing strategy may stop. Hence, this can serve as stopping criteria. However, in cloud-based apps, there could be an infinite number of events that appear in real-time feeding on the cloud. For example, the YouTube smart TV app is presenting new views (i.e., videos) when scrolling down in the application. Practically, there could be an infinite massive number of views. The number of views may also vary with each new start of the application. 

One solution to this challenge is to define a finite number of iteration in which the creeper can iterate over the application or limiting the number of views to be covered before the stop.

\subsection{Test Suite Ripper}

When generating the test cases, we expect some obsolete or invalid test cases. For example, some detected views during the creeping process may not be valid, and still, they may be presented in the test cases. To this end, there is a need for a test ripper to repair those test cases which are not valid. The test ripper may follow an algorithm to repair the test cases. For example, defining several predefined patterns of the invalid test cases or transitions from a view to another view. 

Another repairing process of the test cases could be unique from the remote device. For example, those color buttons on the remote device could be used for several functional and non-functional requirements depending on the application configuration.

\subsection{Test Runner}

When the creeper detects the views, and the test cases are generated and repaired by the test generator and ripper, a test runner is needed to run these test cases. A test runner is merely taking the test suite and run the test cases one by one automatically. Here, the same test runner strategy in android app testing could be followed by the smart TV app testing. However, executing the test cases depends on the development kit. 

\subsection{Fault Taxonomy and Categorization\label{subsec:Fault-Taxonomy}}

After running the test cases on the application, an important task is to identify the encountered faults and the test cases in which these faults related to. However, faults in smart TV app  are not known yet. Here, classical mutation testing is not applicable. For example, recently, Deng \textit{et al.} \cite{Deng2017} have identified different faults in the Android apps within a mutation testing framework for mobile devices. In fact, those faults are more Android-oriented faults, and they are not applicable here. In addition, some of those faults are related to the Activity faults, for example, changing the screen orientation, which is also not appropriate because the Smart TV screen is too big to be frequently oriented. Normally, classical mutation test tools like MuDroid \cite{MoranMDroid2018} or MuJava \cite{Ma:2005:MAC} are used for mobile, web or desktop apps. As we mentioned, those tools are platform-specific tools. An important effort in this approach is done by Cui \textit{et al.} \cite{Cui2017}. Cui \textit{et al.} identified eight different types of faults in smart TV applications. These faults are, TV system halt, TV system reboot, displaying a black screen, having voices but no images, playing images with delaying, application exit by exceptions, playing images with a blurry screen, key has no response, or the response key is wrong. While this is an excellent effort toward the fault categorization, there is a need to identify more faults related to the application itself. Some of those identified faults may also relate to the TV device itself. Also, there is a need to identify a method for how to inject these faults in the smart TV. A significant effort that can be done here is to conduct a study to define the taxonomy of faults in Smart TV apps. A useful input to this study could come from smart TV industry especially those companies which are tracking and getting feedback from users in the cloud. Doing an analytical study on this data to categorize these faults would be an excellent finding.

\subsection{Defining Test Oracle}

Defining the pass and fail criteria is a challenging task in software testing process. Within test automation, the mechanism for determining whether a given test case is passed or failed is named test oracle. In this context, the distinction between the correct and incorrect behavior is called \textquotedblleft test oracle problem\textquotedblright{} \cite{Earl2015}. A classical way to approach the test oracle is the manual identification of the pass and fail by the developer. However, for a significant amount of test cases, this is not accurate and impractical. 

Automating test oracles in smart TV app testing is not an easy task since we don't know precisely the nature and the kind of faults the application face. In addition, the dynamic behavior of the cloud-based smart TV applications may lead to random new views that can be loaded. In fact, this task is connected to the fault taxonomy and categorization discussed in Section \ref{subsec:Fault-Taxonomy}. When we know the faults and can categorize them, we can define the test oracle for the automated testing framework.

\section{Towards an Automated Testing Strategy\label{sec:Towards-a-Systematic-testing}}

\begin{figure*}
\begin{centering}
\includegraphics[scale=0.4]{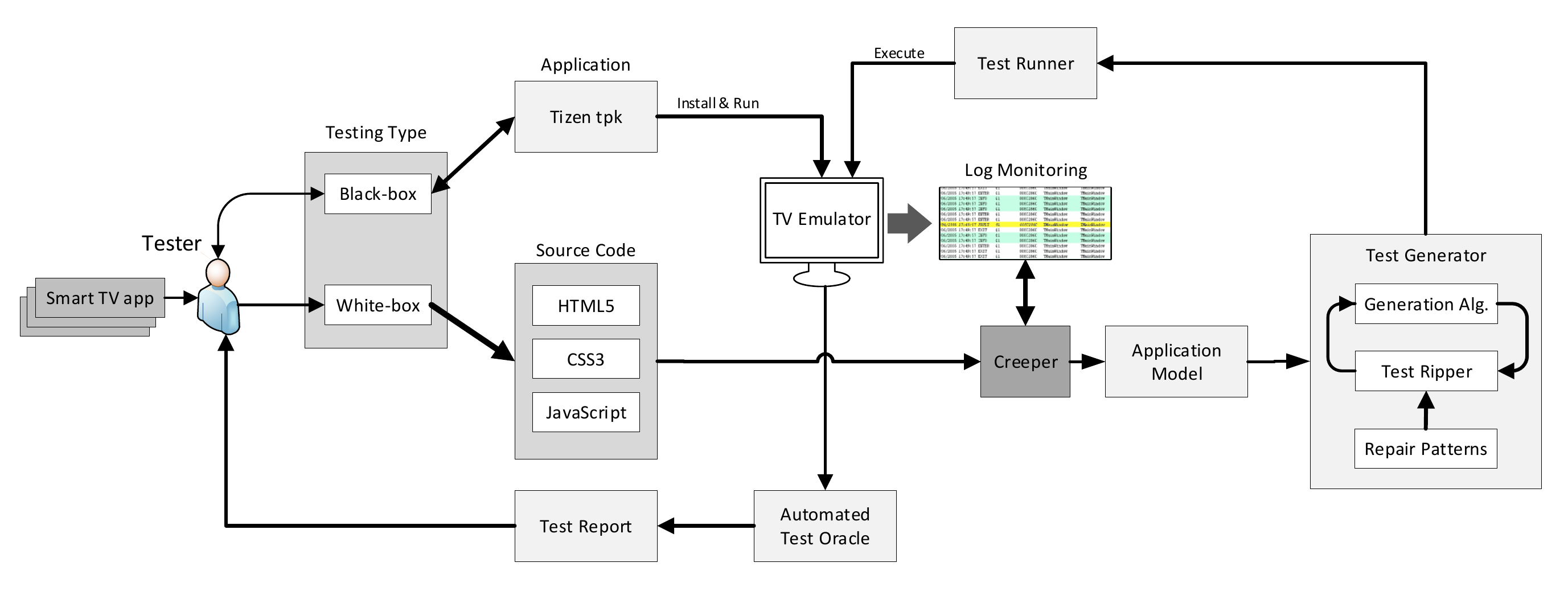}
\par\end{centering}
\caption{\label{fig:Smart-TV-Application-Framework}Smart TV App Testing Framework }
\end{figure*}

Based on the problems and challenges presented so far, here we can propose an automated framework to test the smart TV apps. This framework presents our vision for a strategy to automate the testing process. The framework is working in the Tizen SDK, which includes a smart TV emulator; however, the framework is a general framework and it is applicable for other possible emerging SDKs in the future. Figure \ref{fig:Smart-TV-Application-Framework} shows an overview of this framework and illustrates the essential components and their relationship to each other. 

The framework supports both white and black box testing styles. The tester chooses among these two features depending on the source code availability and the application type. As mentioned previously, even when the source code is available, when the application is a cloud-based app, the tester must consider this case as black-box testing. When the source code is available, the tester will import the project and let the framework do the rest automatically. Here, the creeper will scan the source code and tries to identify the essential views in the UI. 

In case of black-box testing or cloud-based app, which is probably the most critical case, the creeper must use a special algorithm to creep and detect all the views. Detail of this algorithm is presented in the following section (Section \ref{sec:Application-Creeper}). Here, the creeper uses the log messages from the TV emulator to validate the views. 

In both white or black box testing approaches, the creeper will detect the essential views and convert all the views and their relationship with each other to a state machine graph model. This model will be the input to the test generator which consists of a model-based algorithm for generation and also a test Ripper to repair the test cases. The repair will be based on some predefined patterns of invalid test cases. This process is iterative until as far as there is an invalid test case. The framework will execute these test cases through a test runner on the TV emulator, and an automated test oracle module will validate them one by one. Finally, a test report will be presented to the user again.

\subsection{Application Creeper\label{sec:Application-Creeper}}

To detect all the necessary views in the application that need to present in the model for test generation, we have developed an algorithm called EvoCreeper. In fact, object detectors in UI for mobile, desktop, and web apps is not new. There are some algorithms called crawlers to crawl on the UI and detect these objects. None of those algorithms are useful here since we have an entirely different user interaction behavior in the smart TV apps. Besides, we have thought that the name \textquotedblleft creeper\textquotedblright{} suites perfectly with what we want to do as the \textquotedblleft crawler\textquotedblright{} word gives a different meaning due to its use in web and search engine technologies. Algorithm \ref{alg:Application-Creeper} shows the steps of the EvoCreeper.

\begin{algorithm}
\textbf{\small{}Input:}{\small{} $v_{1}$ is the user selected view}{\small \par}

\textbf{\small{}Output:}{\small{} List of views to be modeled $L_{v}$}{\small \par}

Iteration $It\leftarrow$1

Maximum Iteration $It_{max}\leftarrow max$

\textbf{While} $((It<It_{max})\parallel(newView\neq null))$

{\small{}$\quad$ Use $v_{1}$ as a start point}{\small \par}

{\small{}$\quad$ From $v_{1}$ generate five possible directions
$D_{Up}$ , $D_{Down}$ , $D_{Left}$ , $D_{Right}$, $OK$ }{\small \par}

{\small{}$\quad$ }\textbf{\small{}For}{\small{} each direction }{\small \par}

{\small{}$\quad\quad$ Navigate a step}{\small \par}

{\small{}$\quad\quad$ Monitor emulator log for reaction}{\small \par}

{\small{}$\quad\quad$ }\textbf{\small{}If}{\small{} $newView=Active$}{\small \par}

{\small{}$\quad\quad\quad$add $newView$ to $L_{v}$}{\small \par}

{\small{}$\quad\quad$}\textbf{\small{}End If }{\small \par}

{\small{}$\quad$}$It++$

{\small{}$\quad$}\textbf{\small{}End For}{\small \par}

\textbf{End While}

\caption{\label{alg:Application-Creeper} EvoCreeper Steps }
\end{algorithm}

If the focus point is not set by the app developer, the EvoCreeper starts by an action from the tester to choose at least one view to start from, otherwise, it will start from the focused view. From this view, the creeper will start creeping the UI evolutionary and incrementally. The algorithm takes four directions {\small{}$D_{Up}$ , $D_{Down}$ , $D_{Left}$ , $D_{Right}$ } plus the $OK$ button from each view to move. When a new view discovered in each direction (i.e., {\small{}$newView=Active$}), the algorithm will add it to the {\small{}list of views to be modeled $L_{v}$. This algorithm will continue until there are no new discovered views. Here, as another stopping criterion, the algorithm will take some preset number of iteration to avoid the endless discovery loop in some special cases of cloud-based apps. In the following section (Section \ref{subsec:Proof-of-Concept}), we present an example as a graphical proof of concept for this algorithm. }{\small \par}

\subsection{Proof of Concept\label{subsec:Proof-of-Concept}}

\begin{figure*}
\begin{centering}
\includegraphics[scale=0.32]{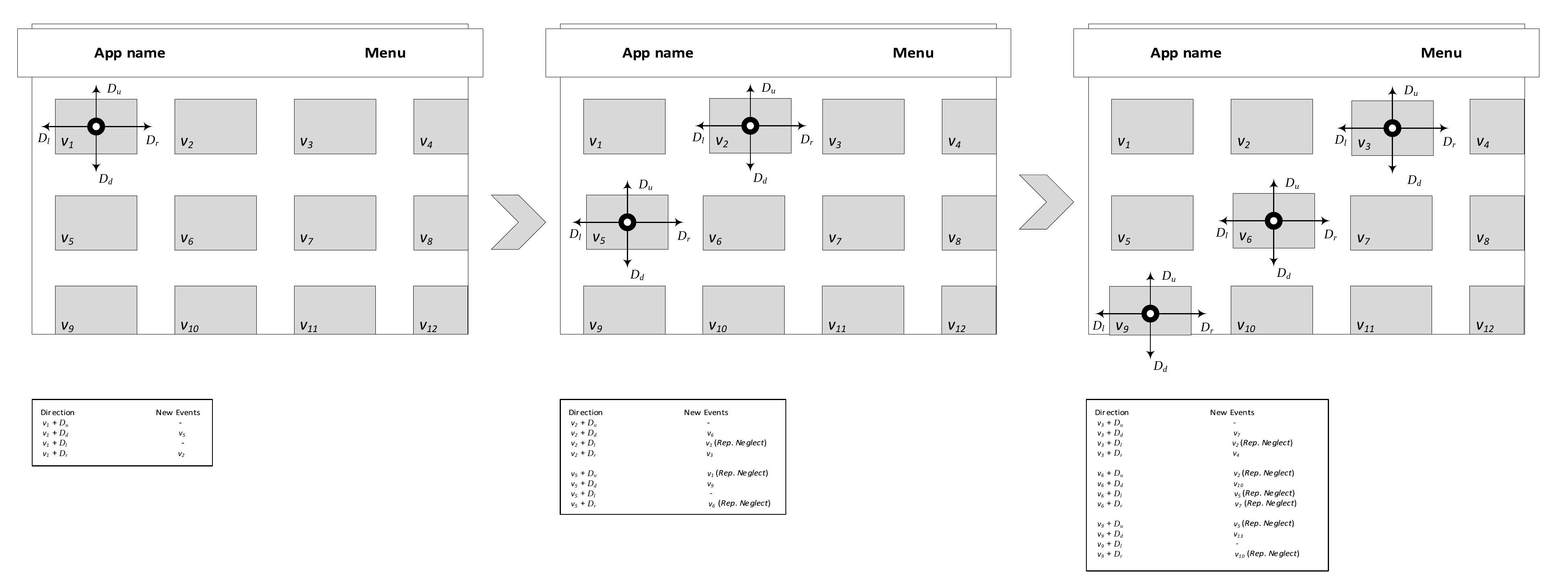}
\par\end{centering}
\caption{\label{fig:Proof-of-concepts}Proof of concepts of the EvoCreeper}
\end{figure*}

In this section, we present a proof of concept for the application creeper in Algorithm \ref{alg:Application-Creeper}. Here, we consider a cloud-based app as a pilot example as it is the most difficult scenario. As shown in Figure \ref{fig:Proof-of-concepts}, each activity window has 12 views and as the user shift down or right, new activities may appear. We consider three iterations of the algorithm. We assume that the tester will choose $v_{1}$ as a start point. In fact, $v_{1}$ is the worst case choice of the views and we observed that choosing the view in the middle of the window may lead to less iteration and better recognition of the views. From $v_{1}$, the algorithm will consider four main directions, {\small{}$D_{Up}$ , $D_{Down}$ , $D_{Left}$ , $D_{Right}$ }{\small \par} plus the $OK$ button. However, here, we will consider only those four directions because the $OK$ button may open a new window in the app. 

For each direction, the creeper algorithm will check for new events, which are most likely new views. Considering the first iteration, and starting from $v_{1}$, the up and left directions {\small{}$D_{u}$, $D_{l}$ will not lead to new views, while the right direction $D_{r}$ leads to }$v_{2}$ and the down direction {\small{}$D_{d}$ leads to }$v_{5}$. For the next iteration, the creeper will start from newly discovered views, $v_{2}$ and $v_{5}$ here. From $v_{2}$, the news views $v_{3}$and $v_{6}$ identified by the creeper algorithm. In addition, $v_{1}$is discovered in the {\small{}$D_{l}$ direction, however, it is neglected by the creeper as it is already available on the view list. Considering the $v_{5}$, the views $v_{1}$, $v_{9}$, and $v_{6}$ are in the three direction $D_{u}$, $D_{d}$, and $D_{r}$ respectively; however, only $v_{9}$ considered as a new view. }{\small \par}

The third iteration also starts from the newly discovered views, {\small{}$v_{3}$, $v_{6}$, and $v_{9}$. In the same way, considering the four directions from each view and filtering all repeated views, four new views were identified, $v_{4}$, $v_{7}$, $v_{10}$, and $v_{13}$. }{\small \par}

The EvoCreeper algorithm works in an iterative evolutionary style to discover new views and events in the application under test. As mentioned, this pilot example considers the cloud-based app. Hence, there is no expectation of the finite numbers of views in the application. To this end, our proposed stopping criteria could be useful here. The creeper algorithm will continue for a certain number of iterations or when no new views discovered.

\section{Functional and Non-functional Testing Opportunities in Smart TV Applications}\label{FunctionaltestingSection}

For testing the functional or non-functional requirement in smart TV, we need a measure. This measure can be used in the test generation process as a coverage criterion and also can be used in the design of test oracle. While for functional requirement it is straightforward, converting a non-functional requirement into an exact measure is a tricky task. Here, an approximation could be useful. 

Many problems could be addressed here. For example, addressing the minimum hardware requirements for a specific smart TV application would be an interesting idea to investigate. Most of the smart TV devices nowadays in the market rely on low computation power CPU and memory. Extra hardware may be used to measure the energy consumption of the CPU during the testing process. 

Covering the event interactions in different level is also interesting functional testing. Here, full, partial, or systematic coverage of the events is the decision that must be made by the tester. Also, a comparison of these three coverage criteria is an important study topic to know which approach is better for fault finding. 

The limitation in memory and CPU lead to another interesting non-functional requirement that may also be used in the testing process, which is the execution time. It would be interesting to know the situation and sequences in the smart TV application that causes long or short execution time. This could also be useful to identify and detecting security vulnerabilities. In fact, security is an essential issue in smart TV applications that have never been addressed before.

Probably, an essential non-functional requirement that must be addressed in smart TV applications is the usability. Due to the availability of remote device, the usability testing is necessary. In fact, the remote device remains the main constraint facing the usability of the smart TV applications. At this early research stage, it is useful to address how to make the applications more usable and what are the factors that affect the usability. It is true that the user-oriented testing technique could be more realistic here; however, an automated testing method could support the final result of usability testing report.

\section{Conclusion and Future Work\label{sec:Conclusion}}

In this paper, we have presented the key ingredients, challenges, and some proposed solutions for the smart TV app testing. We think that in the near future, smart TV apps will be an essential piece of software in the whole context of IoT services. Despite this importance, we can't find a systematic and robust testing strategy in the literature for the smart TV apps. After an extensive study of these applications, we discover many open problems and challenges in which we illustrated them in this paper. We found that the most crucial problem to be solved is the test generation strategy. In this paper, we proposed a fully automated framework to test smart TV apps. In addition, we have also illustrated our EvoCreeper algorithm that creeps the views available in the application window. The algorithm uses an iterative evolutionary style to discover new views. The output of the algorithm will be input to the test generator strategy that generates the necessary test cases for the automated testing framework. 

Depending on the testing process, there are many opportunities for smart TV app testing. For example, the security, usability, scalability, and robustness testing are essential issues that have not been addressed in the literature. Here, our proposed framework is also useful for these non-functional properties by just altering the test oracle and test generator components.As part of our work, we are planning to present more comprehensive strategy with testing results of different smart TV apps in the future.

\section*{Acknowledgment}

This research is conducted as a part of the project TACR TH02010296
Quality Assurance System for Internet of Things Technology.

\bibliographystyle{unsrt}
\bibliography{SmartTVPaperFile}

\end{document}